\documentclass[preprint]{elsarticle}
\usepackage[utf8]{inputenc}
\usepackage{hyperref}
\usepackage{lineno}
\modulolinenumbers[5]

\usepackage{amsmath}

\def\useBW{0}

\usepackage{siunitx}
\usepackage{upgreek}
\usepackage[version=4]{mhchem} %chemical formula gas

\usepackage{booktabs} %tables

\journal{Nuclear Instruments and Methods in Physics Research Section A}

\begin{document}

\begin{frontmatter}

\title{Performance of a GridPix detector based on the Timepix3 chip}


\author[Nikhef]{C. Ligtenberg\corref{correspondingauthor}}
\cortext[correspondingauthor]{Corresponding author. Telephone: +31 617 377 014}
\ead{cligtenb@nikhef.nl}

\author[Nikhef,Bonn]{K. Heijhoff}
\author[Bonn]{Y. Bilevych}
\author[Bonn]{K. Desch}
\author[Nikhef]{H. van der Graaf}
\author[Nikhef]{F. Hartjes}
\author[Bonn]{J. Kaminski}
\author[Nikhef]{P.M. Kluit}
\author[Nikhef]{G. Raven}
\author[Bonn]{T. Schiffer}
\author[Nikhef]{J. Timmermans}

\address[Nikhef]{Nikhef, Science Park 105, 1098 XG Amsterdam, The Netherlands}
\address[Bonn]{Physikalisches Institut, University of Bonn, Nussallee 12, 53115 Bonn, Germany}

\begin{abstract}
A GridPix readout for a TPC based on the Timepix3 chip is developed for future applications at a linear collider. 
The GridPix detector consists of a gaseous drift volume read out by a single Timepix3 chip with an integrated amplification grid. 
Its performance is studied in a test beam with 2.5 GeV electrons. 
The GridPix detector detects single ionization electrons with high efficiency. 
The Timepix3 chip allowed for high sample rates and time walk corrections. 
Diffusion is found to be the dominating error on the track position measurement both in the pixel plane and in the drift direction, and systematic distortions in the pixel plane are below \SI{10}{\um}. 
Using a truncated sum, an energy loss (dE/dx) resolution of 4.1\% is found for an effective track length of \SI{1}{m}.
\end{abstract}

\begin{keyword}
Micromegas\sep gaseous pixel detector\sep micro-pattern gaseous detector\sep Timepix\sep GridPix
\end{keyword}

\end{frontmatter}

%\linenumbers

\section{Introduction}
In the context of a Time Projection Chamber for a future linear collider a gaseous pixel detector  is developed based on the Timepix3 chip. The GridPix single chip detector discussed here, allows for a detection of single electrons with a granularity of \SI{256 x 256}{} pixels with a pitch of \SI{55 x55}{\um}. By counting the number of pixel hits, the number of clusters can be estimated allowing for a precise measurement of the energy loss dE/dx. 

Since the invention of the device \cite{Colas:2004ks,Campbell:2004ib}, a series of developments have taken place that culminated in GridPix detectors using the Timepix chip \cite{Kaminski:2017bgj}. In this paper the first results using a Timepix3 based Gridpix detector will be described. In the design of the detector, special attention has been given to minimize the distortions in the pixel plane and drift direction in order to meet the tracking precision needed for a TPC at a linear collider. The device can also be applied for medical imaging, proton radiotherapy or used in other particle physics experiments \cite{Krieger:2017yax}. Here test beam results taken at the ELSA test beam facility in Bonn will be presented. Some results using this device in a laser setup were presented at TIPP17 \cite{TIPP17}.

\section{Description of the GridPix device}
A GridPix is a CMOS pixel readout chip with a gas amplification grid added by photolitho\-graphic postprocessing techniques \cite{Kaminski:2017bgj}.
%A GridPix is a gaseous pixel detector \cite{Kaminski:2017bgj}. 
In this work, the GridPix consists of a Timepix3 chip \cite{Poikela:2014joi} with a \SI{4}{\um} thick Silicon-rich Nitride protective layer, and \SI{50}{\um} high SU8 pillars that support the \SI{1}{\um} thick Al grid that has \SI{35}{\um} diameter circular holes aligned to the pixels. The growing of the protection layer of Timepix chips has been further optimized at the Fraunhofer Institute for Reliability and Microintegration (IZM) in Berlin, making the device more spark proof. An ionizing particle will liberate electrons in the TPC drift volume that will drift  towards the grid and enter the avalanche region. The avalanche yields an electronic signal on the pixel. The Timepix3 chip has low noise ($\approx$70 e$^-$) and allows per pixel for a  precise measurement of the Time of Arrival (ToA) and the Time over Threshold (ToT) using a TDC (clock frequency \SI{640}{MHz}). For the readout the SPIDR software is used \cite{Visser:2015bsa}. 

\begin{figure}
    \centering
    \includegraphics[width=0.45\textwidth]{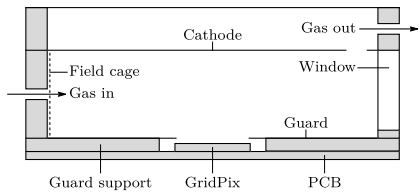}
    \caption{Schematic drawing of the GridPix detector.}
    \label{fig:crossSection}
\end{figure}

In \autoref{fig:crossSection} a cross section of the GridPix detector\linebreak(\SI{14.1 x 14.1}{mm}) located in a small drift volume is shown. The box has a length of \SI{69}{mm}, a width (not shown) of \SI{42}{mm} and a height of \SI{28}{mm} with a maximum drift length of about \SI{20}{mm}. The beam enters the drift volume through the \SI{5}{mm} thick synthetic window from the right side. The electric drift field is defined by a series of parallel conductive strips in the cage and is set to about \SI{280}{V/cm}. On the guard plane - located \SI{1}{mm} above the grid - a voltage is applied that matches the local drift voltage.

\section{Test beam measurement}

\begin{figure}
    \centering
    \includegraphics[width=0.6\textwidth]{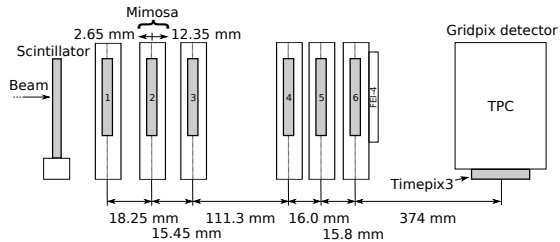}
    \caption{Setup with telescope and GridPix detector.}
    \label{fig:setup}
\end{figure}

In July 2017, measurements were performed at the ELSA test beam facility in Bonn. 
The ELSA storage ring delivered a beam of \SI{2.5}{GeV} electrons at a rate set to a maximum of \SI{10}{kHz}. Each beam cycle of approximately \SI{6.5}{s} contained one spill of about \SI{5}{s}. 
To acquire a precise reference track, a silicon pixel tracking telescope was introduced in the setup as shown in \autoref{fig:setup}.
Electrons from the beam first passed through a scintillator that was used to provide a trigger signal.
This was followed by the tracking Mimosa telescope, consisting of 6 silicon detection planes mounted on a slider stage with each \SI{1152 x 576}{} pixels sized \SI{18.4 x 18.4}{\um}.
Finally, the beam crosses the gas volume of the GridPix detector. The whole GridPix detector was mounted on a remote controllable rotation stage. The beam was perpendicular to the drift direction and at a 17 degrees angle with the chip edges.
On the last telescope plane an inactive FEI4 chip \cite{GarciaSciveres:2011zz} was present that caused multiple scattering of the beam corresponding to a r.m.s. of \SI{0.25}{mm} at the GridPix detector. 

Both the Mimosa telescope and the Timepix3 chip were operated in data driven mode. For synchronization, triggers were numbered by a Trigger Logic Unit (TLU) \cite{TLU} and saved in the two data streams. The Mimosa chips were continuously read out with a rolling shutter taking \SI{115.2}{\us}, meaning that a single frame can contain multiple triggers. 
The Timepix3 hits are attributed to a single trigger by considering all hits within \SI{400}{ns} of a trigger.

During data taking the gas volume of the GridPix detector was flushed with a premixed gas consisting of \SI{95}{\percent} \ce{Ar}, \SI{3}{\percent} \ce{CF4}, and \SI{2}{\percent} \ce{iC4H10}.
This gas - called \emph{T2K TPC gas} - is suitable for a large TPC because of the low diffusion in a magnetic field and its relatively large drift velocity.
The cathode and guard voltage of the GridPix were set such that the electric field was \SI{280}{V/cm}, near the value at which the drift velocity is maximal for this gas. With Magboltz the drift velocity is predicted to be \SI{78.86\pm0.01}{\um/ns} \cite{Biagi:1999nwa}.
To achieve a high efficiency, the grid voltage was set at \SI{350}{V}.
The threshold per pixel was put at 800 e$^-$ to reduce the number of noise hits to a minimum.
The temperature and pressure at time of data taking were stable at \SI{301.6}{K} and \SI{1034.20}{mbar}. 
The Oxygen concentration in the gas was \SI{211}{ppm}. %.1\pm1?
In \autoref{tab:run} the parameters of the analyzed run are summarized.

\begin{table}[]
\caption{Parameters of the analyzed run. The error on temperature and pressure indicates the spread during the run.}
\label{tab:run}
    \centering
    \begin{tabular}{l r}
\toprule
%\multicolumn{2}{c}{Run parameters}\\\midrule
Run duration & 60 minutes\\
Triggers & 4 733 381 \\
$V_\text{grid}$ & 350 V \\
$E_\text{drift}$ & 280 V/cm \\
Rotation (around $z$-axis) & 17 degree \\
Rotation (around $y$-axis) & 0 degree \\
Threshold & 800 e$^-$ \\
Temperature & \SI{301.63\pm0.08}{K}\\
Pressure & \SI{1034.20\pm0.05}{mbar}\\ %?
Oxygen concentration & \SI{211}{ppm} \\ \bottomrule
    \end{tabular}
\end{table}

From the measured ToA of the Timepix3 hits, the $z$-position is calculated using the predicted drift velocity of \SI{78.86}{\um/ns}. This value is found to be consistent, but because of systematic uncertainties there was no attempt at a precise determination.

\section{Track reconstruction and event selection}

\subsection{Track fitting}
%In order to find candidate hits for a track, simple cluster finding is applied. Hits from the GridPix detector (telescope) are binned by their position perpendicular to the beam in a histogram of \si{12 x 12} (\si{30 x 15}) bins. If more than 6 (3) hits in one bin, the bin is combined with its 8 neighbours. A cluster is found if 10 (5) or more hits are found.

To reconstruct a track, a straight line is fitted to the hits. 
The $x$-axis is chosen parallel to the beam, and the drift direction is parallel to the $z$-axis. Apart from a rotation, the pixel column and row coordinates correspond to $x$ and $y$, respectively.
Tracks are fitted using a linear regression fit in $y(x)$ and $z(x)$.
Hits are assigned errors in the 2 directions perpendicular to the beam $\sigma_y$, $\sigma_z$. This will be discussed in detail in section \ref{sec:resolutiony} and \ref{sec:resolutionz}.

To achieve an accurate reconstruction of the tracks, the telescope and the GridPix detector have to be aligned. In a first step, the positions of the 6 telescope planes are independently aligned. The planes are placed perpendicular to the beam, and their position along the beam is measured. The 5 rotations and \SI{4 x 2}{} shifts are iteratively determined from data.
In the second step, the GridPix detector is aligned to the beam by rotating it along 3 axes and measuring the shifts in the directions perpendicular to the beam.

\begin{figure}
    \centering
    \if\useBW1
        \includegraphics[width=0.6\textwidth]{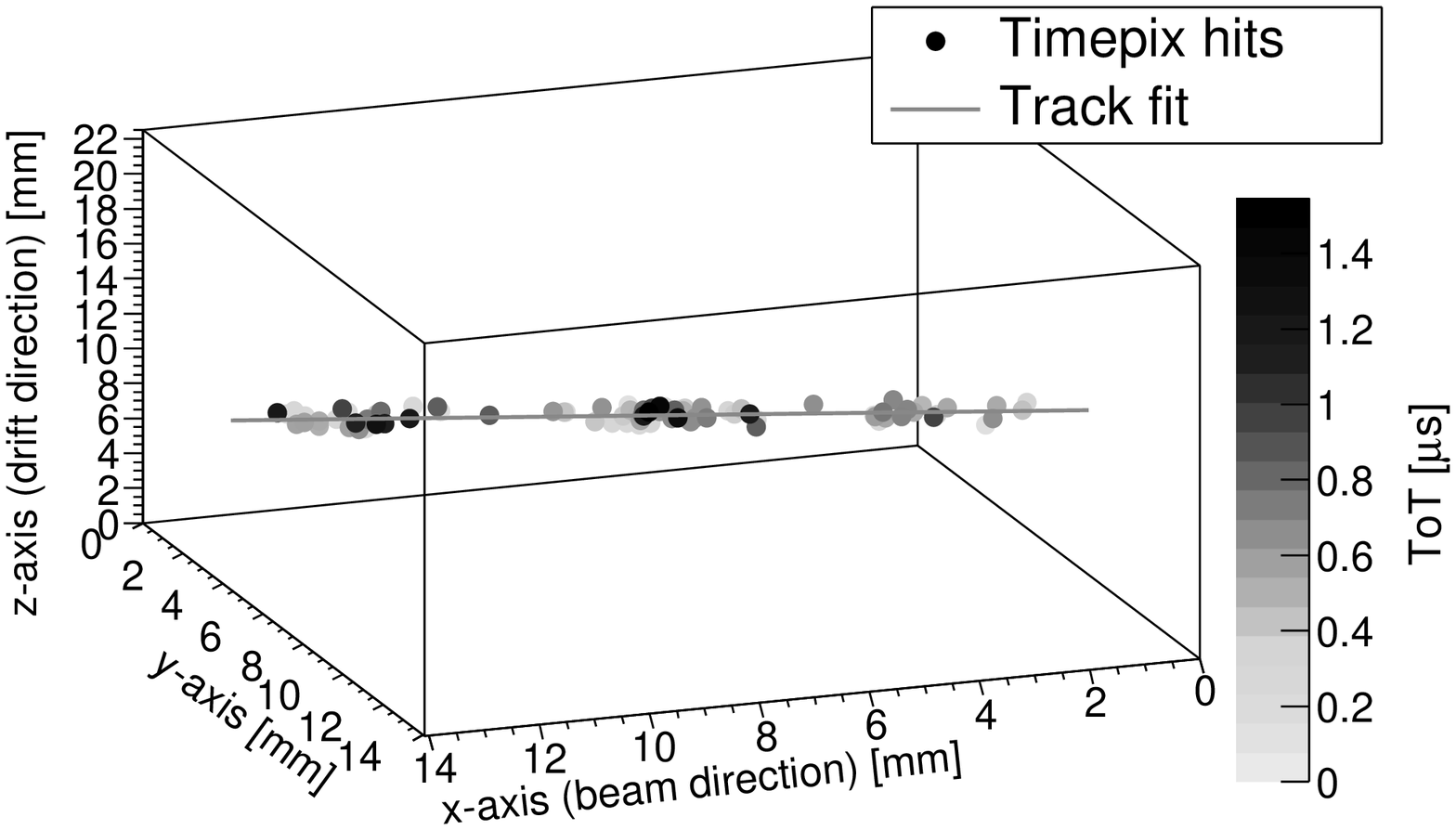}
    \else 
        \includegraphics[width=0.6\textwidth]{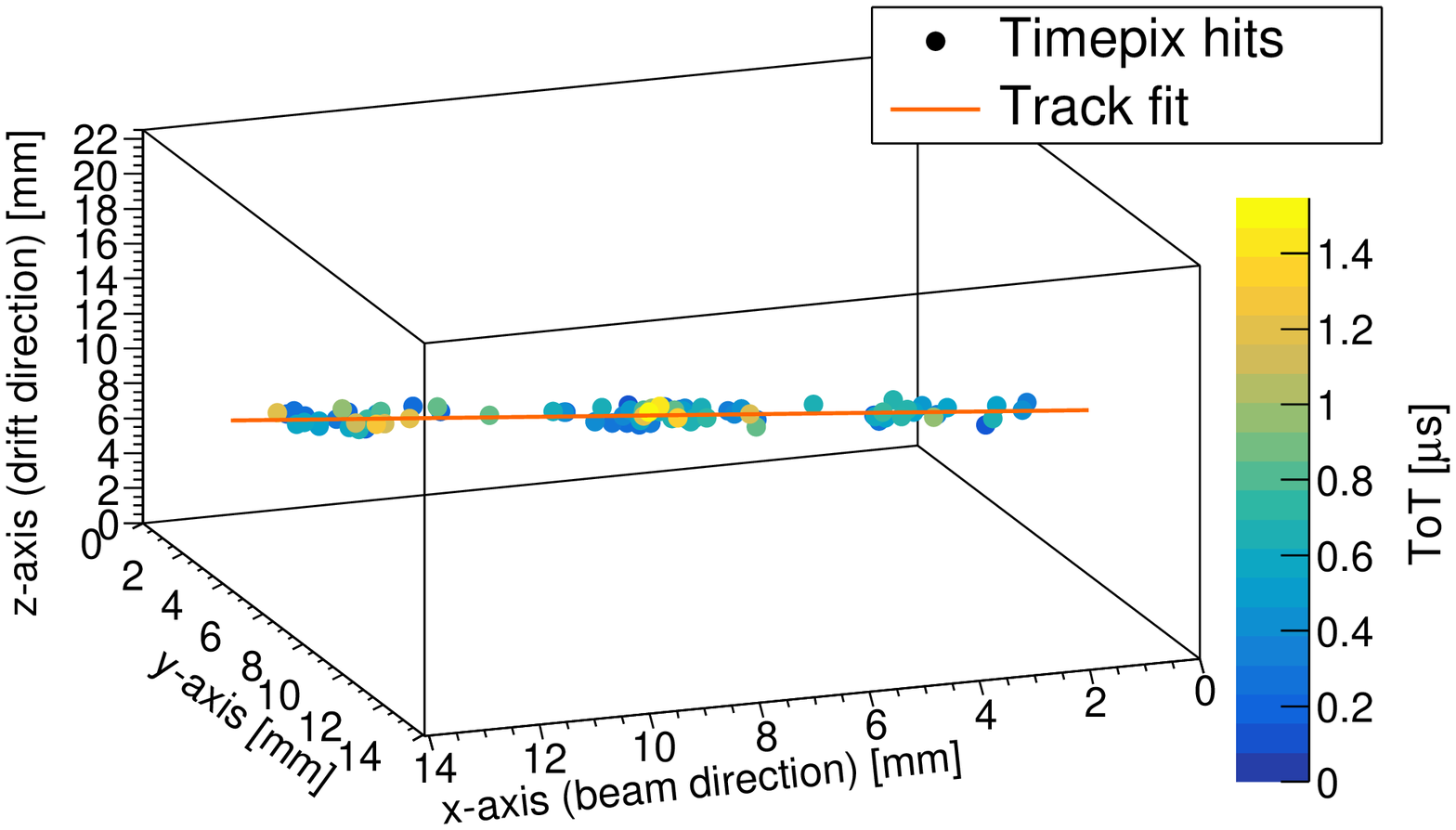}
    \fi
    \caption{An example event with 108 GridPix detector hits including the time walk correction and the track fit.}
    \label{fig:display}
\end{figure}

Since the telescope track is affected by multiple scattering, the most precise track fit is obtained by fitting the hits from the GridPix detector with the combined hits in the telescope. The hits in the telescope planes are merged in one super-point with a \SI{10}{\um} error at the center of the last telescope plane. An example of GridPix hits with a fitted track is shown in \autoref{fig:display}.

\subsection{Selections}
The performance of the detector is measured using events with one clean track in the GridPix detector and the telescope. Given the large amount of data collected, priority in the selection has been given to clean tracks over efficiency.

In the telescope we require the track to have hits in at least 4 out of the 6 planes. Moreover, the extrapolated telescope track should pass through the active volume of the TPC.
For the GridPix detector we select hits that have at least a magnitude corresponding with a ToT of \SI{0.15}{\us} to reject the hits with the worst time walk error, see section \ref{sec:timewalk}. A track is rejected if it has less than 30 GridPix hits. The GridPix track should pass the whole TPC, that is the projection crosses the first and last pixel column. After a first fit, the refit accepts only hits that are within $3\sigma_{\text{drift}}$ and $2\sigma_{\text{plane}}$. Backgrounds and tracks with delta electrons are suppressed by requiring that at least 75\% of all GridPix hits are used in the track fit and only one track is found.
A telescope-GridPix track pair is defined as matched if the extrapolated telescope track is less than \SI{1}{mm} away from the center of the GridPix track. 
Events with an unmatched track pair or multiple matches (due to the rolling shutter) are rejected.
 
About \SI{69}{\percent} of the events pass all selection cuts. An overview of the selections is given in \autoref{tab:selection}.

\begin{table}
\caption{Table with selection cuts.}
\label{tab:selection}
\centering
\begin{tabular}{l} 
\toprule \multicolumn{1}{c}{Telescope}\\
\midrule
At least 4 planes hit \\
Reject outliers ($>\SI{700}{\um}$) \\ 
Telescope track goes through TPC \\
\hline\multicolumn{1}{c}{GridPix detector} \\
\hline
Hit ToT $>\SI{0.15}{\us}$ \\
At least 30 hits\\
Reject outliers ($>3\sigma_{drift}$, $>2\sigma_{plane}$)\\
At least 75\% of total number of GridPix hits in fit \\
Track projection crosses first and last pixel column \\
\hline\multicolumn{1}{c}{Matching of telescope and GridPix detector}\\
\hline
Tracks closer than \SI{1}{mm} at center of TPC \\
A unique track pair match \\
%\hline\multicolumn{1}{c}{Delta rejection} \\
%\hline
\bottomrule
\end{tabular}
\end{table}

\section{Test beam results}

\subsection{Number of hits on track}
\begin{figure}
    \centering
    \includegraphics[width=0.6\textwidth]{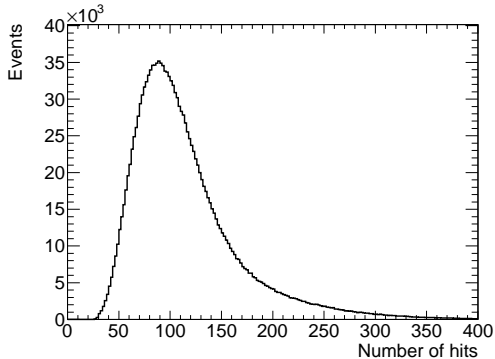}
    \caption{Distribution of number of hits on the track at a grid voltage of \SI{350}{V}.}
    \label{fig:hitDistribution}
\end{figure}

In \autoref{fig:hitDistribution} the number of GridPix hits associated to the track in the fiducial volume (216 pixels) is shown for a grid voltage of 350 V. The most probable number of hits is 91 and the mean is 114 for an effective track length of \SI{12}{mm}. This is in agreement with the expected 100 electron-ion pairs/cm \cite{Patrignani:2016xqp}. The shape of the distribution is Landau-like with a long tail due to delta electrons.

\begin{figure}
    \centering
    \includegraphics[width=0.6\textwidth]{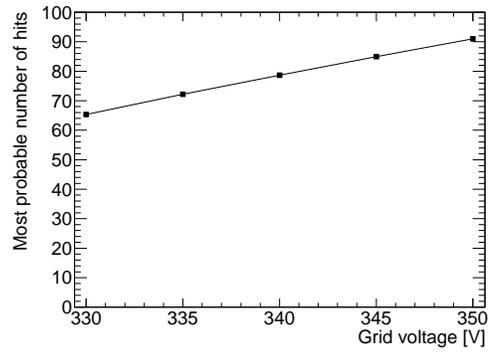}
    \caption{Most probable number of hits on the track as a function of grid voltage.}
    \label{fig:nHits}
\end{figure}

In \autoref{fig:nHits} the most probable number of hits is shown as a function of the grid voltage. 
One expects that the efficiency of the GridPix detector increases with grid voltage until it reaches a plateau at an efficiency of almost \SI{100}{\percent}. Increasing further the grid voltage will induce crosstalk and far above 400 V sparks would be produced. 
The data analyzed in the following was collected at a voltage of \SI{350}{V}, close to the expected plateau and at a high efficiency.
A search for double hits did not yield any indication for crosstalk.

\subsection{Time walk correction}\label{sec:timewalk}
\begin{figure}
    \centering
    \includegraphics[width=0.6\textwidth]{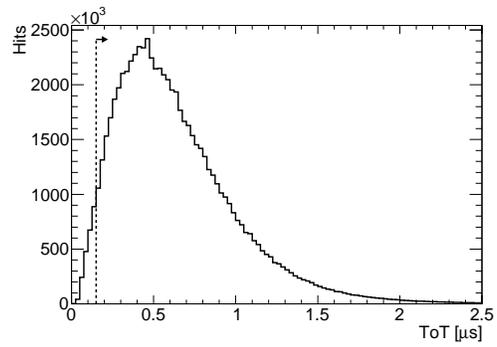}
    \caption{Distribution of the ToT. The arrow indicates the cut on hits with a ToT below \SI{0.15}{\us} }
    \label{fig:ToT}
\end{figure}

A pixel is hit when the collected charge is above the threshold. Because it takes longer to reach the threshold for a small signal than it does for a large signal, the measured ToA depends on the magnitude of the signal. This is called the time walk. The capability to record simultaneously both ToA and ToT per pixel is one of the improvements of the Timepix3 chip over its predecessor the Timepix chip. In \autoref{fig:ToT} the ToT distribution is shown. The Timepix3 chip allows to correct for the time walk by using the ToT as a measure of the signal magnitude. 

%At this point we switch from the telescope frame to the GridPix detector frame. The $x$ and $y$ axes are chosen to match the column and row axes and the $z$-axes corresponds with the drift direction.

\begin{figure}
    \centering
    \if\useBW1
        \includegraphics[width=0.6\textwidth]{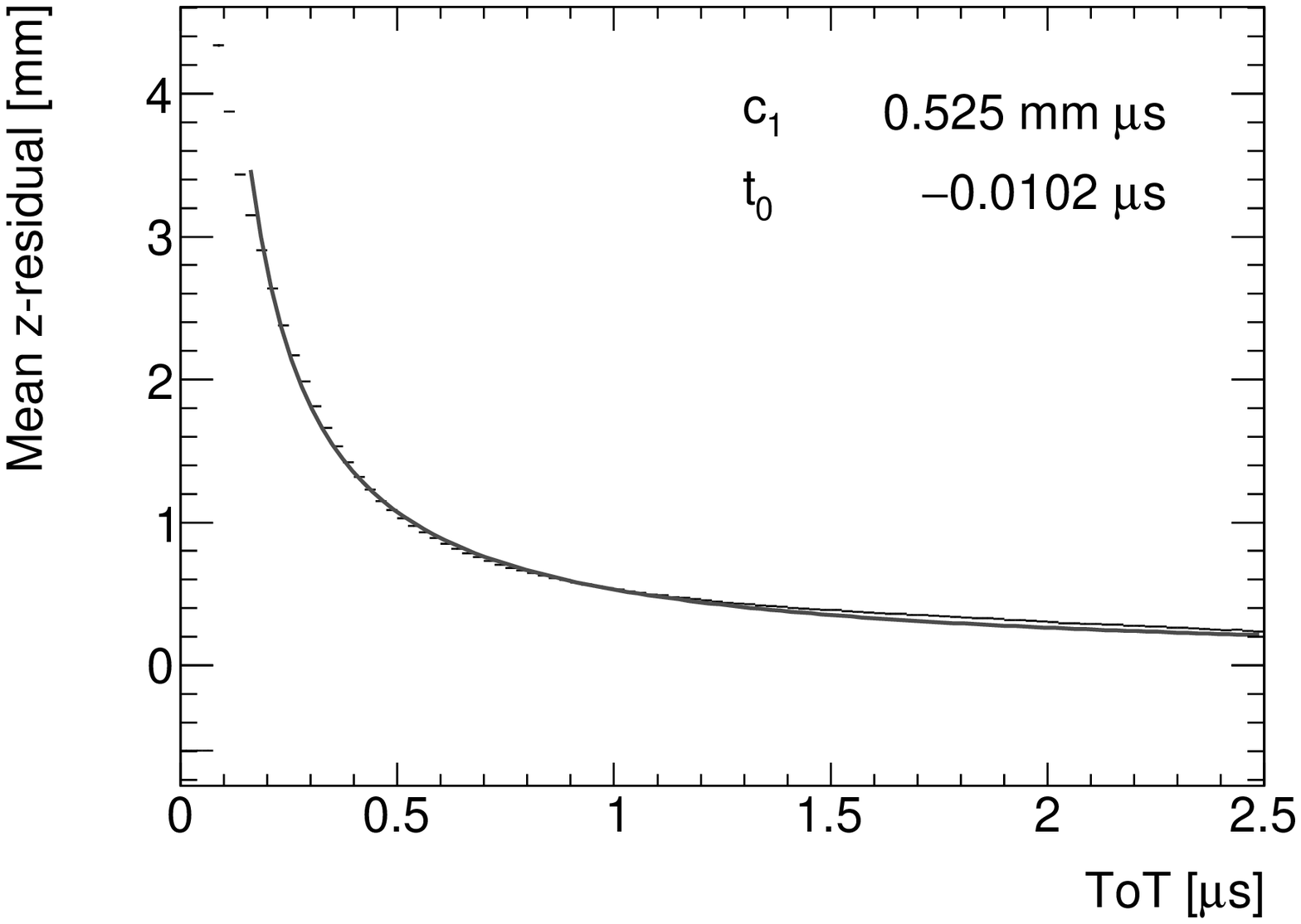}
    \else 
        \includegraphics[width=0.6\textwidth]{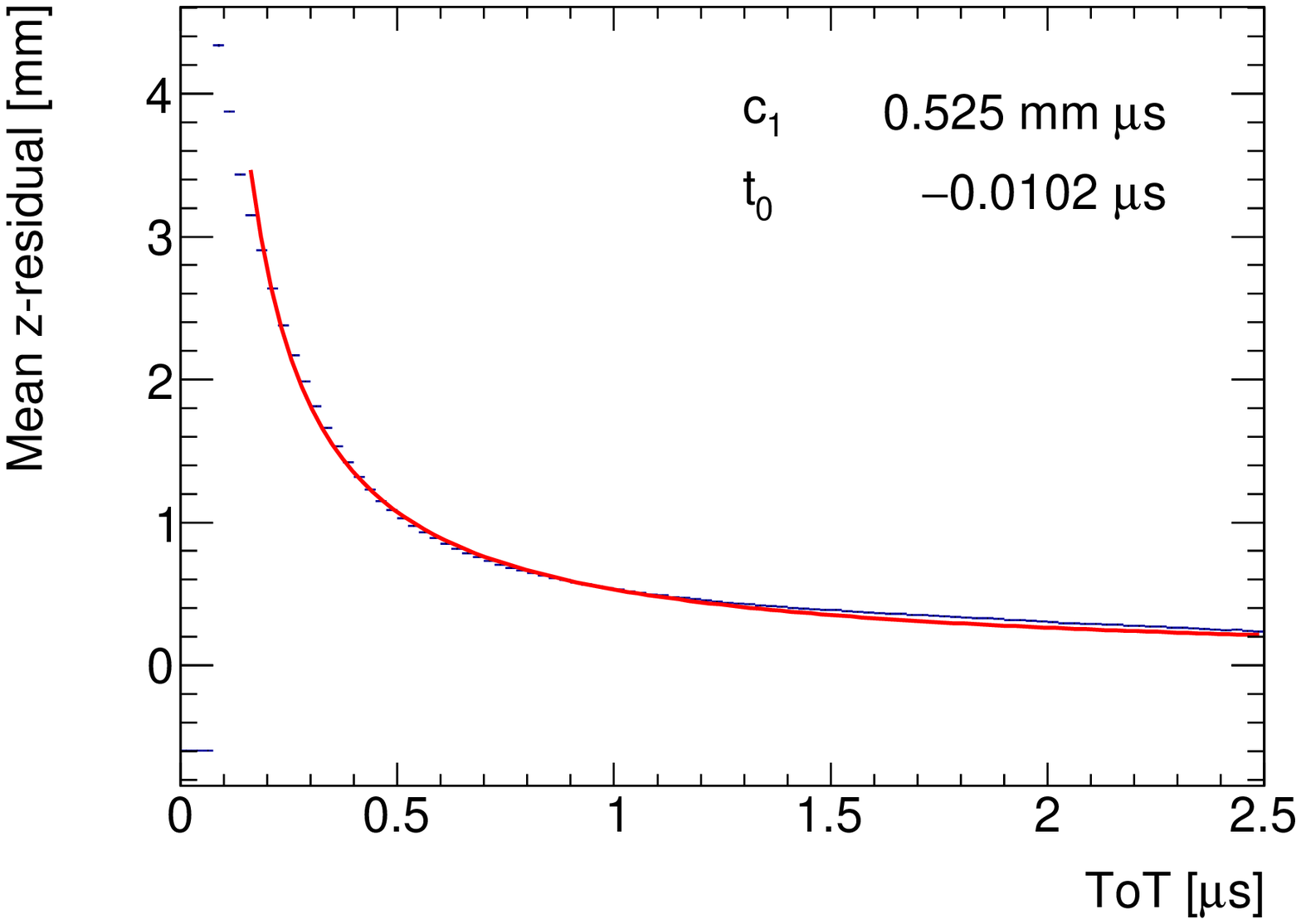}
    \fi
    \caption{Mean $z$-residual without time walk correction as function of ToT, fitted with equation \eqref{eq:timewalk}.}
    \label{fig:timewalkToT}
\end{figure}

\begin{figure}
    \centering
    \if\useBW1
        \includegraphics[width=0.6\textwidth]{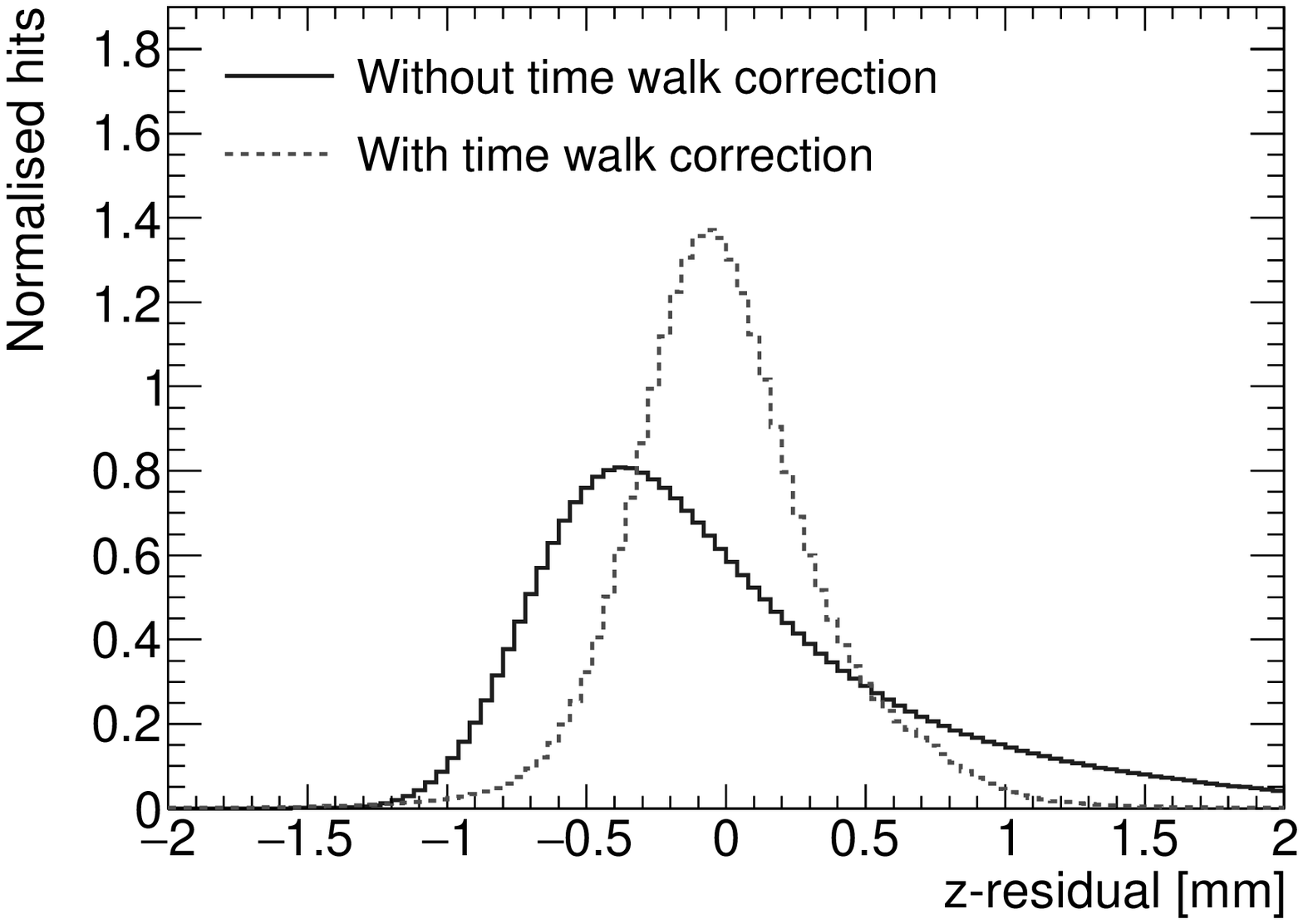}
    \else 
        \includegraphics[width=0.6\textwidth]{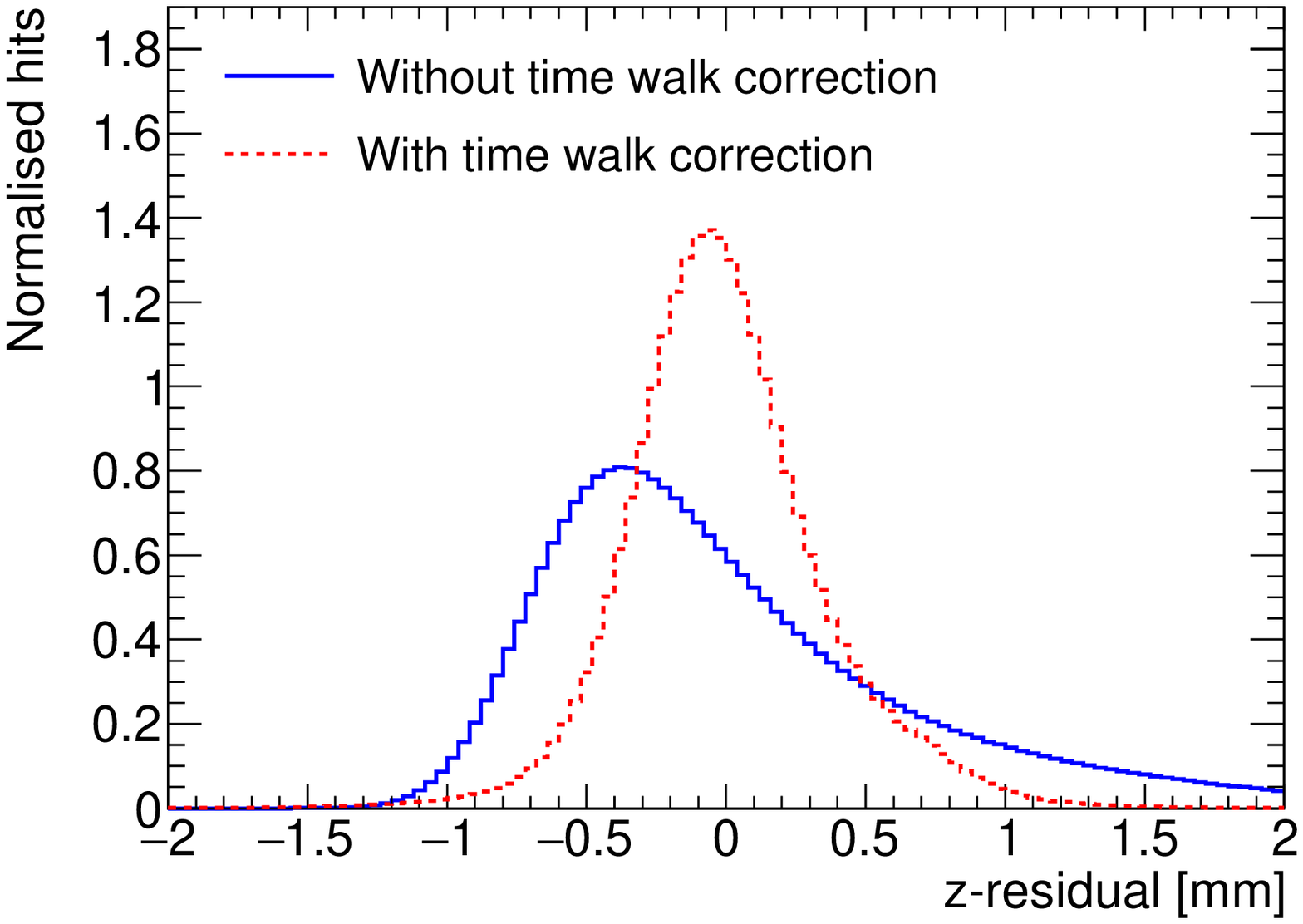}
    \fi
    \caption{Distribution of $z$-residuals before and after time walk correction}
    \label{fig:TWCorrection}
\end{figure}

First, the ToT was found to vary as a function of the column number and therefore per column a correction factor for the ToT was introduced. The uncorrected mean of $z$-residuals - defined as the difference of the track fit prediction and the $z$-position of the hit - is shown as a function of the corrected ToT in \autoref{fig:timewalkToT}. The relation can be parametrized using the time walk $\delta z_\text{tw}$ as a function of the corrected ToT $t_\text{ToT}$:
\begin{equation}
    \delta z_\text{tw} = \frac{c_1}{t_\text{ToT} + t_0},
    \label{eq:timewalk}
\end{equation}
where $c_1$ and $t_0$ are constants determined from a fit. The distribution of $z$-residuals before and after applying the time walk correction, is shown in \autoref{fig:TWCorrection}. Functions with more parameters were also tried, but did not improve the results.

\subsection{Hit resolution in the pixel plane}\label{sec:resolutiony}
\begin{figure}
    \centering
    \if\useBW1
        \includegraphics[width=0.6\textwidth]{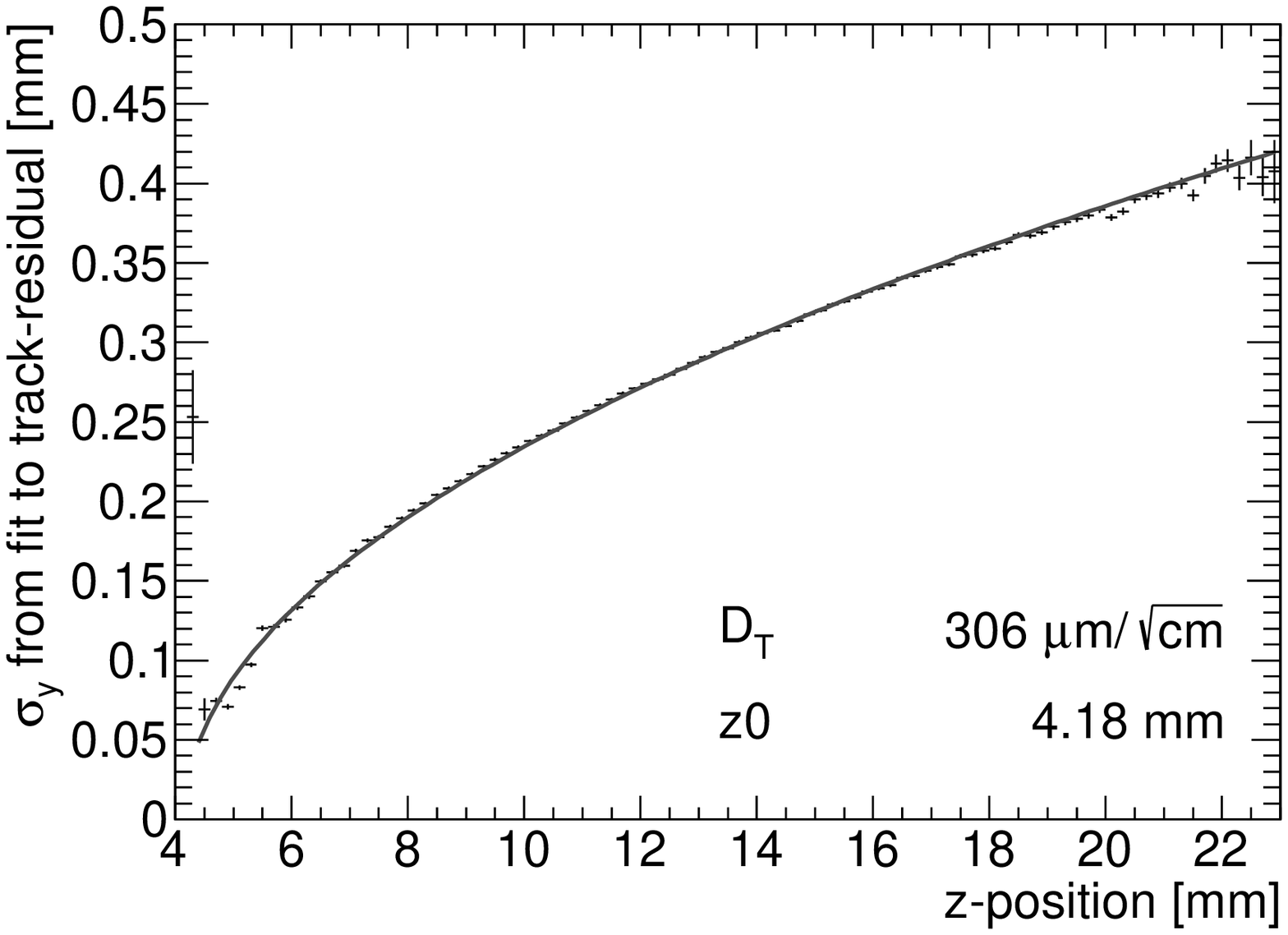}
    \else 
        \includegraphics[width=0.6\textwidth]{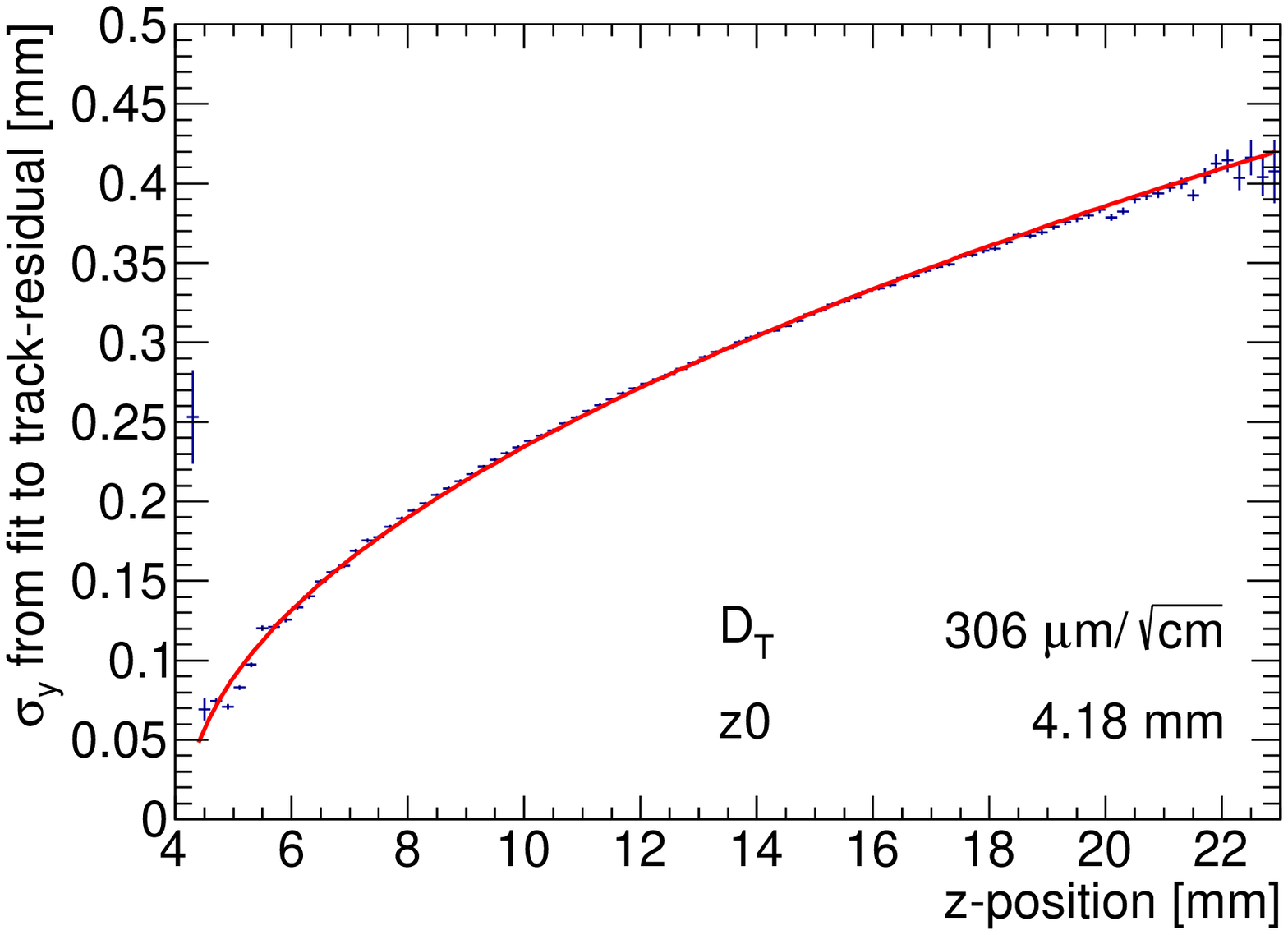}
    \fi
    \caption{Measured hit resolution in pixel plane (blue points) fitted with the resolution according to equation \eqref{eq:sigmay} (red line), where the hit resolution at zero drift distance $d_\text{pixel}/\sqrt{12}$ was fixed to \SI{15.9}{\um}.}
    \label{fig:Diffy}
\end{figure}

The momentum resolution of a TPC depends on the hit resolution in the pixel plane. There are two important contributions to the hit resolution in the pixel plane: a constant contribution caused by the pixel size $d_\text{pixel}$ and a transverse drift component that scales with drift distance and the diffusion coefficient $D_T$. The resolution $\sigma_y$ is given by:
\begin{equation}
    \sigma_y^2=\frac{d_\text{pixel}^2}{12} +D_T^2(z-z_0),
    \label{eq:sigmay}
\end{equation}
where $z_0$ is the position of the grid. The measured hit resolution as a function of $z$-position is shown in \autoref{fig:Diffy}. For each point an estimated systematic uncertainty of \SI{1}{\um} is added to the statistical uncertainty. The diffusion gives the largest contribution to the hit resolution in most of the detector volume. 
The measured diffusion coefficient $D_T=\SI{306}{\um/\sqrt{cm}}$ is lower than the expected $D_T=\SI{318\pm7}{\um/\sqrt{cm}}$.

\subsection{Hit resolution in the drift direction}\label{sec:resolutionz}
\begin{figure}
    \centering
    \if\useBW1
        \includegraphics[width=0.6\textwidth]{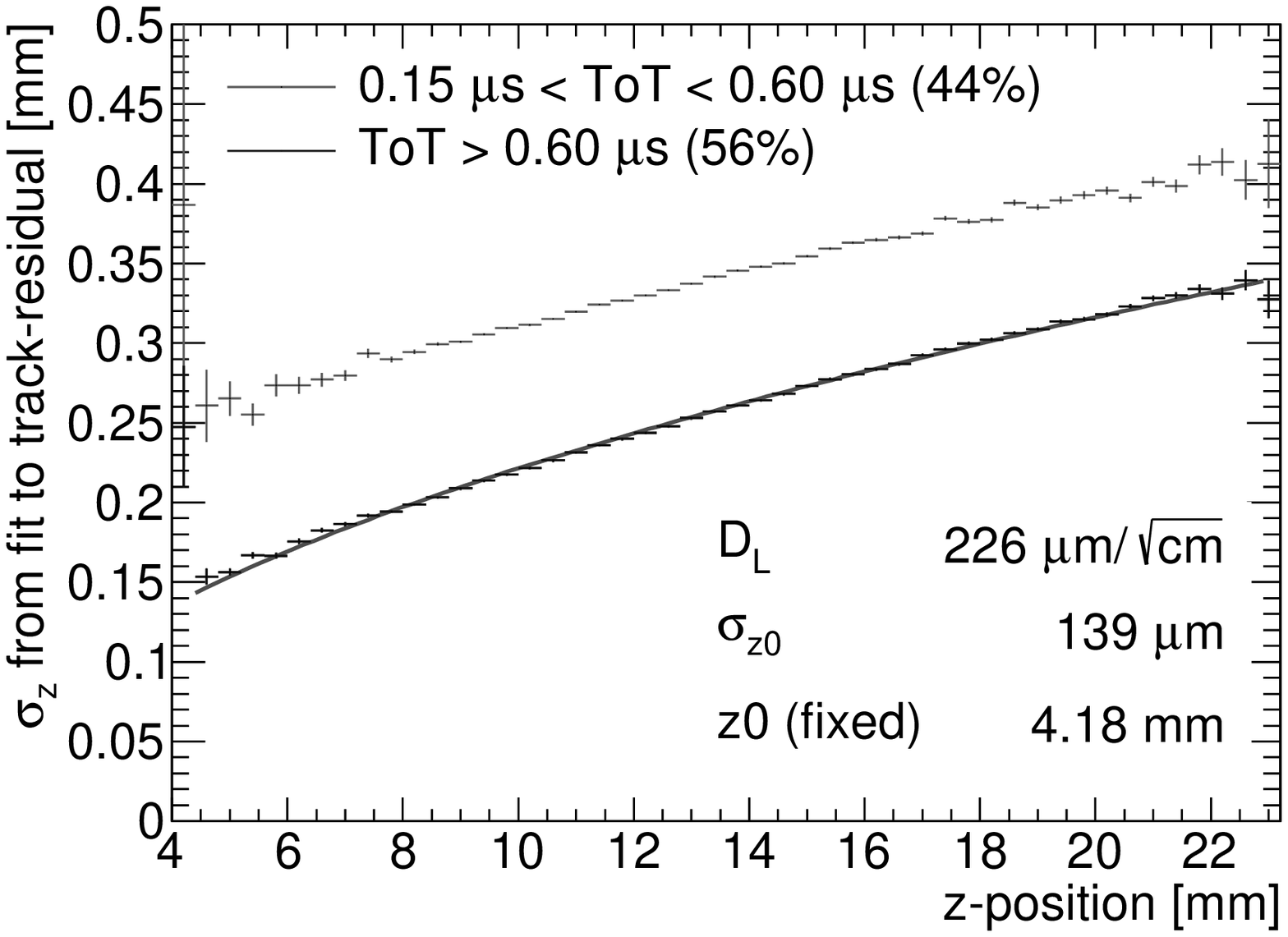}
    \else 
        \includegraphics[width=0.6\textwidth]{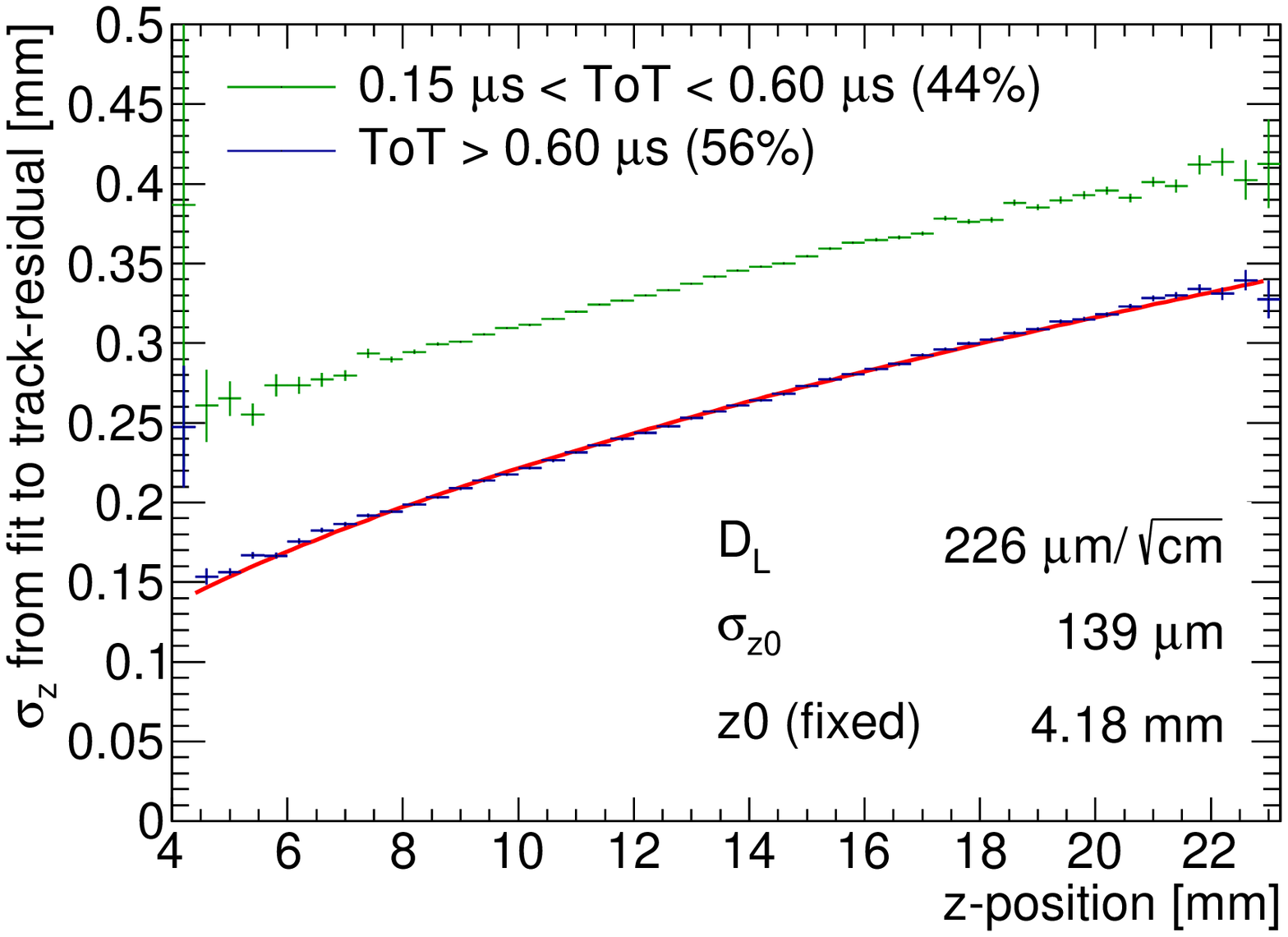}
    \fi
    \caption{Measured hit resolution in drift direction split by ToT. The hits with a ToT above \SI{0.60}{\us} (blue points) are fitted with the resolution according to equation \eqref{eq:sigmaz} (red line). In the legend the fraction of hits in both selections is given.}
    \label{fig:Diffz}
\end{figure}

The hit resolution in the drift direction is related to the ToA distribution. There are three contributions. A constant contribution from the time resolution $\tau=\SI{1.56}{ns}$, a contribution from other noise sources such as jitter and time walk, and a contribution from longitudinal diffusion with coefficient $D_L$. The resolution $\sigma_z$ is given by
\begin{equation}
    \sigma_z^2=\frac{\tau^2 v_\text{drift}^2 }{12}+\sigma_{z0}^2+D_L^2(z-z_0)
    \label{eq:sigmaz}
\end{equation}
The error from time walk is largest for small ToT, motivating a split of the hits roughly in half at a ToT of \SI{0.60}{\us}. The hit resolution in the drift direction is shown in \autoref{fig:Diffz}. To each point an estimated systematic error of \SI{1}{\um} was added. The grid position was fixed to $z_0=\SI{4.18}{mm}$ found in the fit to \autoref{fig:Diffy}. The diffusion is found to be $D_L=\SI{226}{\um/\sqrt{cm}}$, which is higher than the expected value of $D_L=\SI{201\pm5}{\um/\sqrt{cm}}$.

\subsection{Deformations}
\begin{figure}
    \centering
    \if\useBW1
        \includegraphics[width=0.6\textwidth]{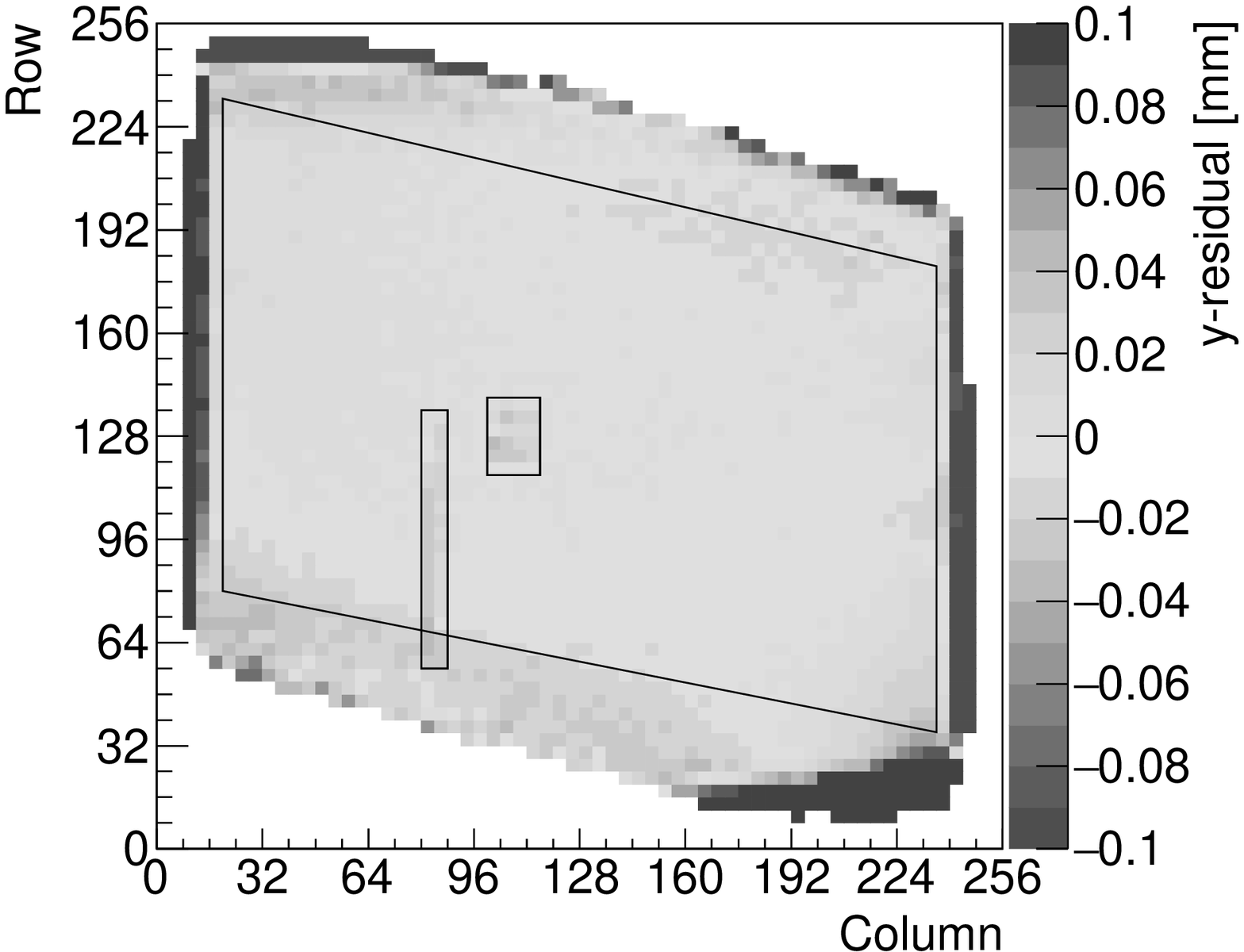}
    \else 
        \includegraphics[width=0.6\textwidth]{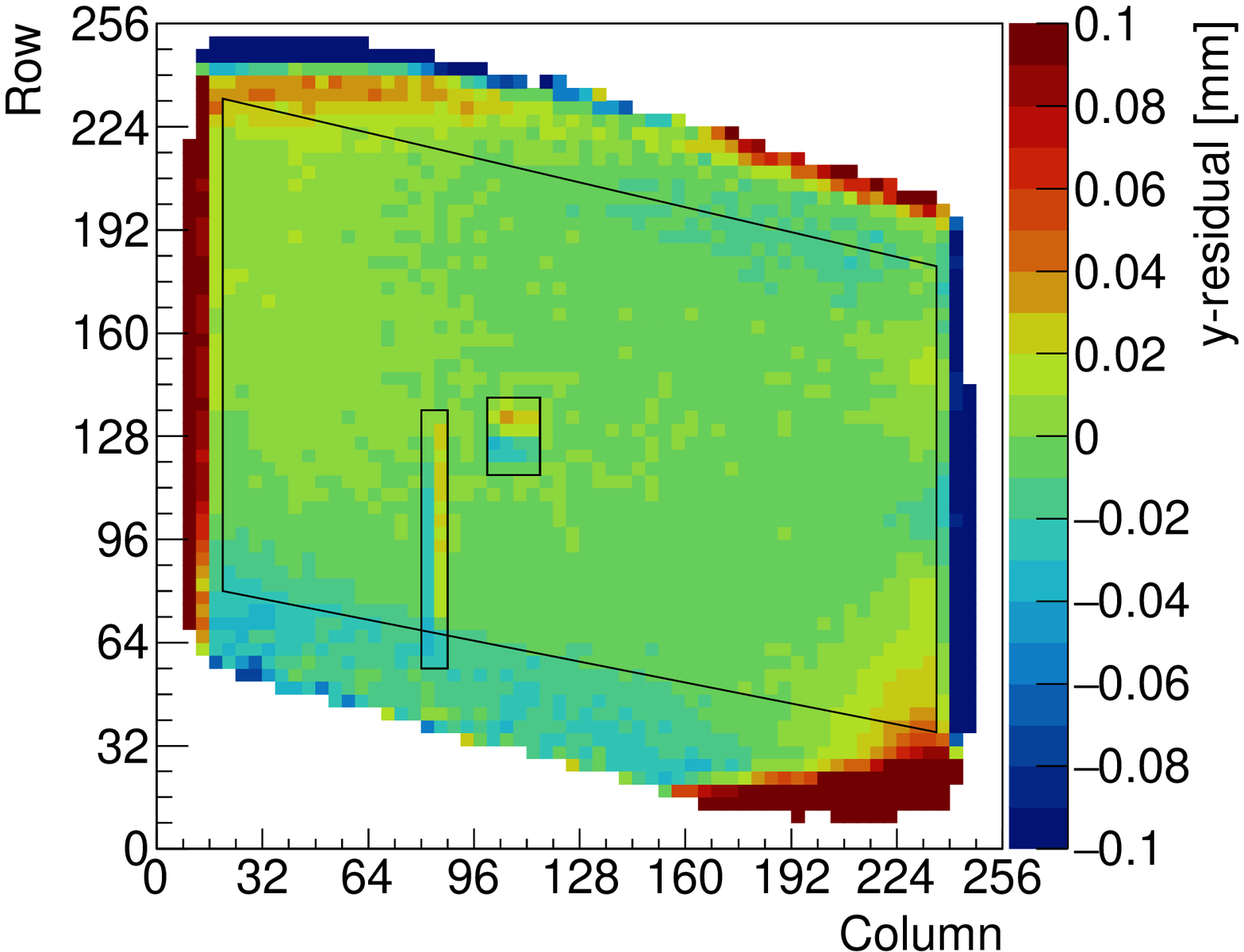}
    \fi
    \caption{Mean residuals in the pixel plane at the expected hit position.}
    \label{fig:deformyexp}
\end{figure}

\begin{figure}
    \centering
    \if\useBW1
        \includegraphics[width=0.6\textwidth]{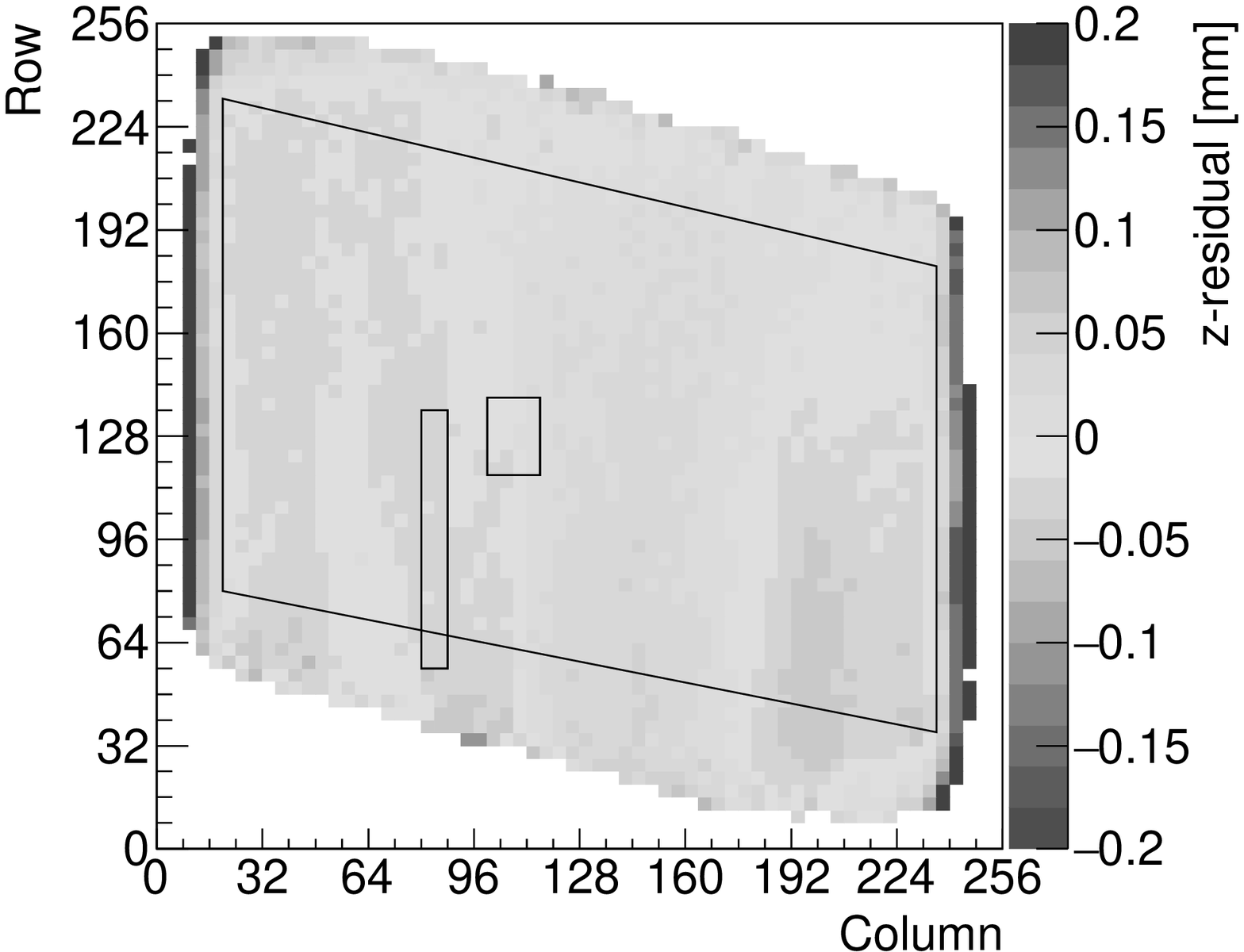}
    \else 
        \includegraphics[width=0.6\textwidth]{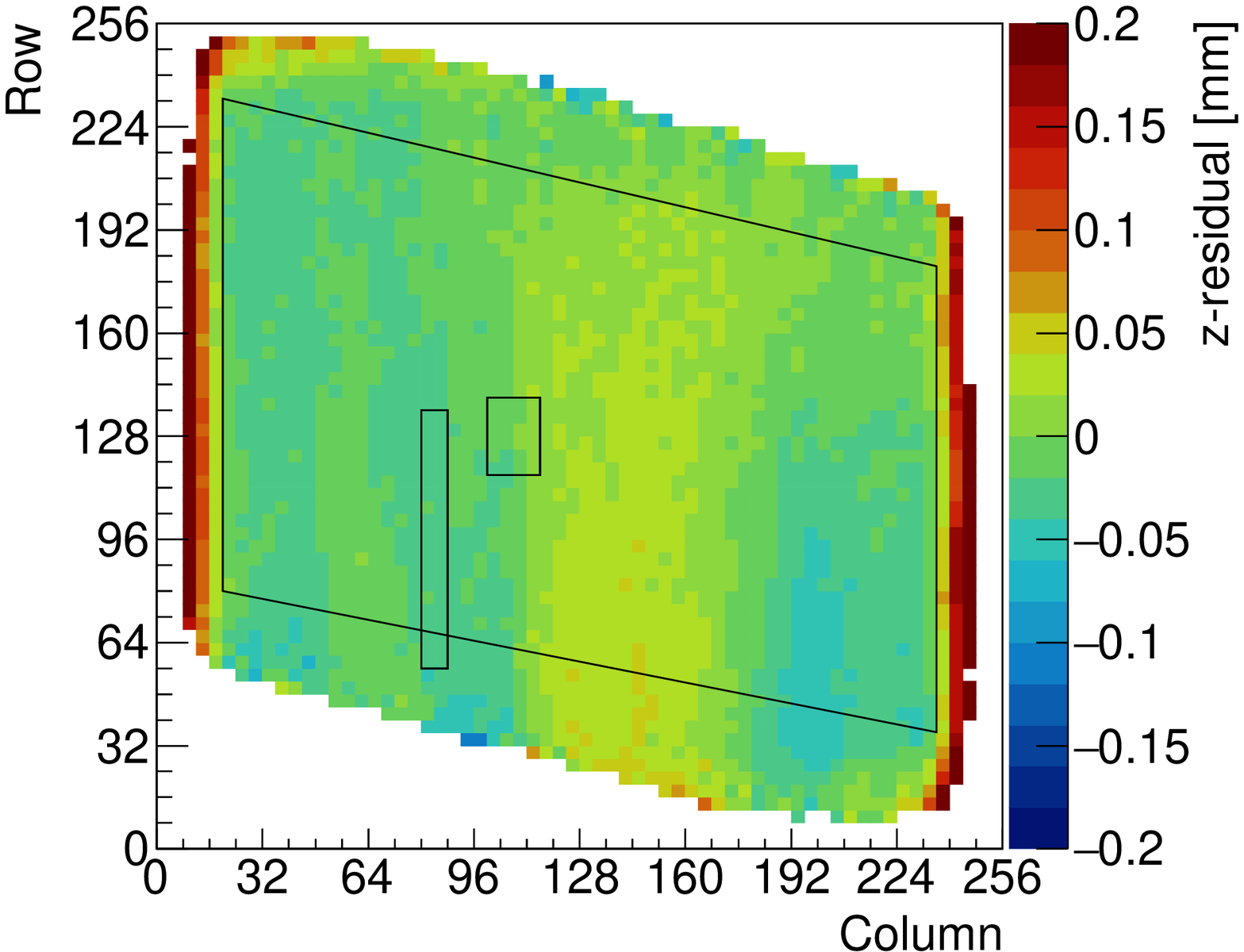}
    \fi
    \caption{Mean residuals in the drift direction at the expected hit position.}
    \label{fig:deformxexp}
\end{figure}

\begin{figure}
    \centering
    \if\useBW1
        \includegraphics[width=0.6\textwidth]{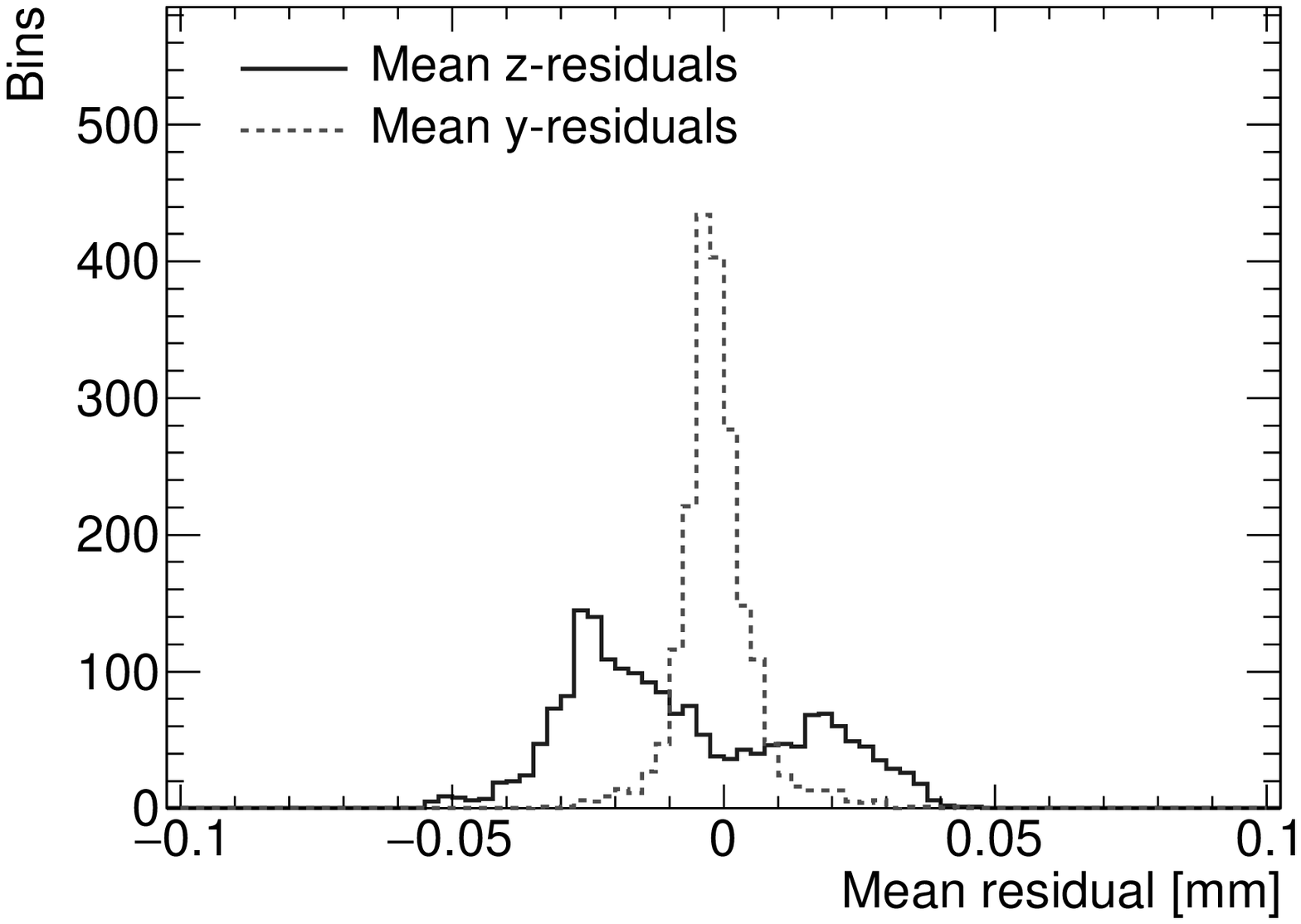}
    \else 
        \includegraphics[width=0.6\textwidth]{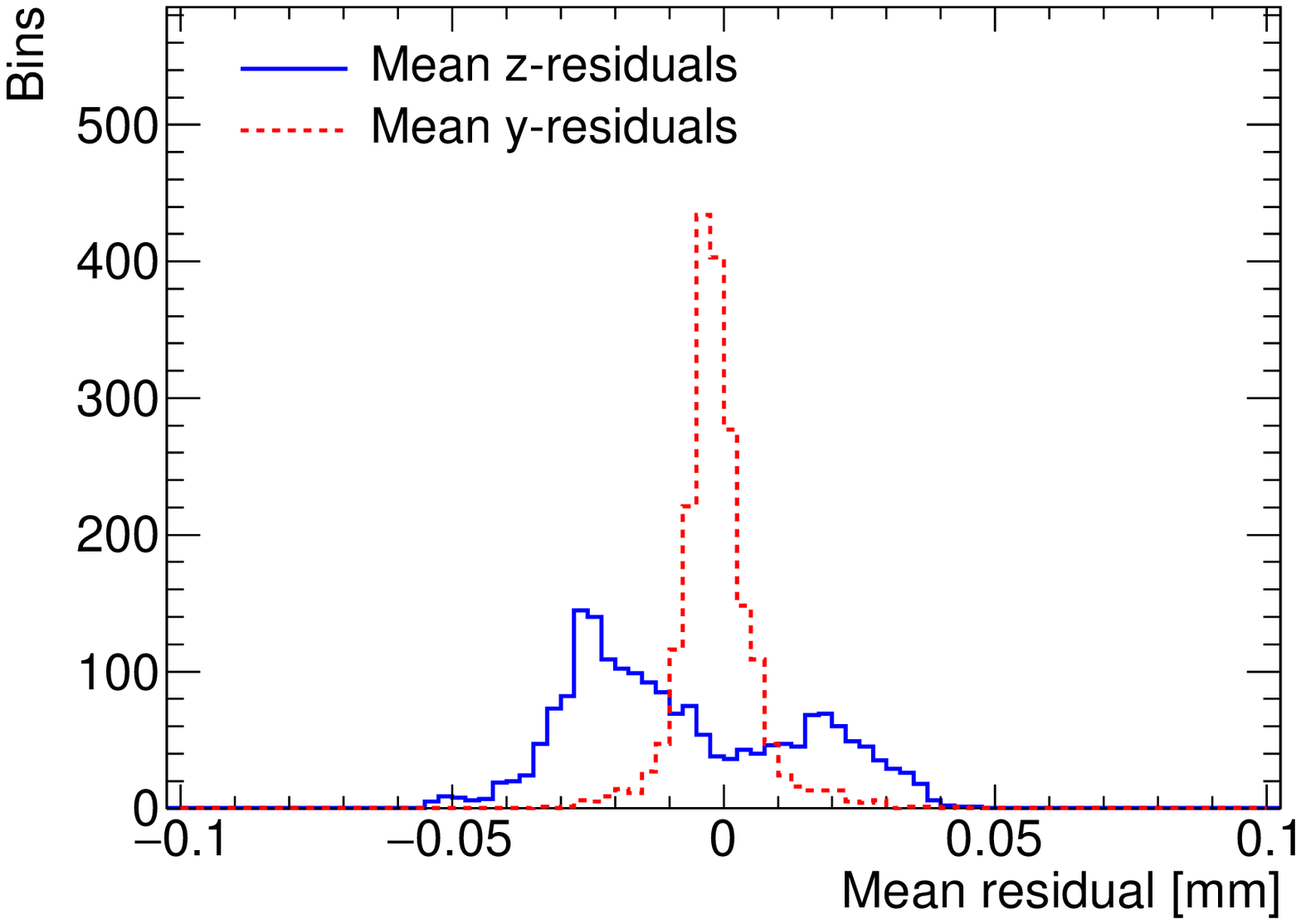}
    \fi
    \caption{Distribution of the mean residuals from \SI{4 x 4}{} bins within the selected region.}
    \label{fig:deformFreq}
\end{figure}

For a large TPC with pixel readout it is important that systematic deviations of position measurements are small and stay well below typically \SI{20}{\um}. Here we study deformations in the pixel plane and the drift direction, caused by for example distortions in the drift field or geometric uncertainties. The chip is divided in \SI{64 x 64}{} bins of \SI{4 x 4}{} pixels each for which the mean deformation is calculated. For every hit, the expected originating position on the track is traced and the residual is filled at that bin. In \autoref{fig:deformyexp} and \autoref{fig:deformxexp} the mean residual in the $xy$-plane and the mean $z$-residual are shown, respectively. In the diagram only bins with more than 1000 entries are shown. Only bins in the selected fiducial area were used to make the distribution of the mean residuals shown in \autoref{fig:deformFreq}. The r.m.s. of the deviations is \SI{7}{\um} in the pixel plane and \SI{21}{\um} (\SI{0.3}{ns}) in the drift direction, respectively. This means that the systematics on the position measurement in the pixel plane - the bending plane of the TPC - are less than \SI{10}{\um}.


\subsection{Energy loss measurement}
A TPC can identify particles using their characteristic energy loss. The GridPix detector measures the energy loss dE/dx by counting the number of detected electrons. Because of the large fluctuation in energy loss, the mean is dominated by a few high energy deposits. To retrieve a better estimate, the truncated sum of electrons is calculated.

\begin{figure}
    \centering
    \if\useBW1
        \includegraphics[width=0.6\textwidth]{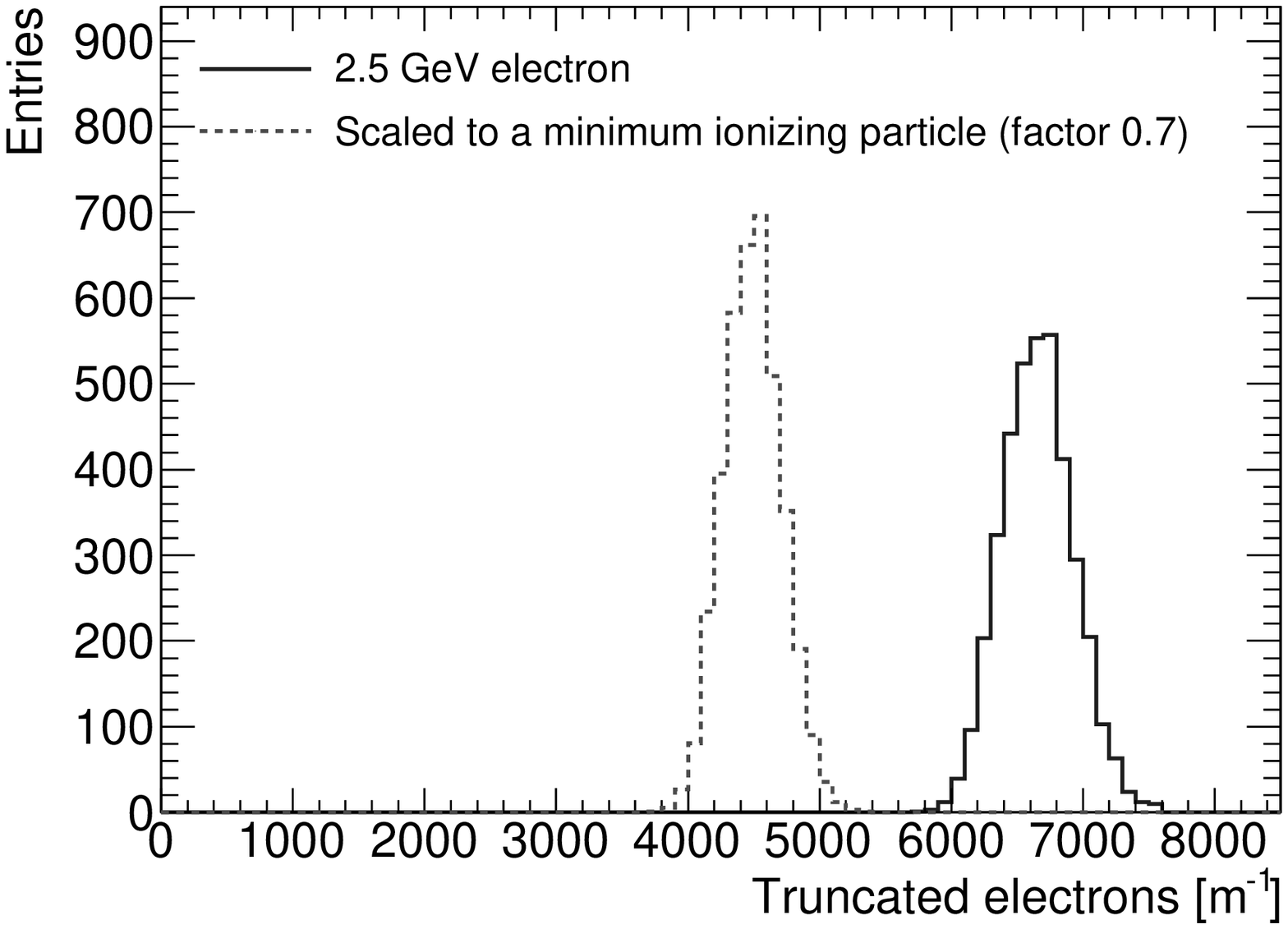}
    \else 
        \includegraphics[width=0.6\textwidth]{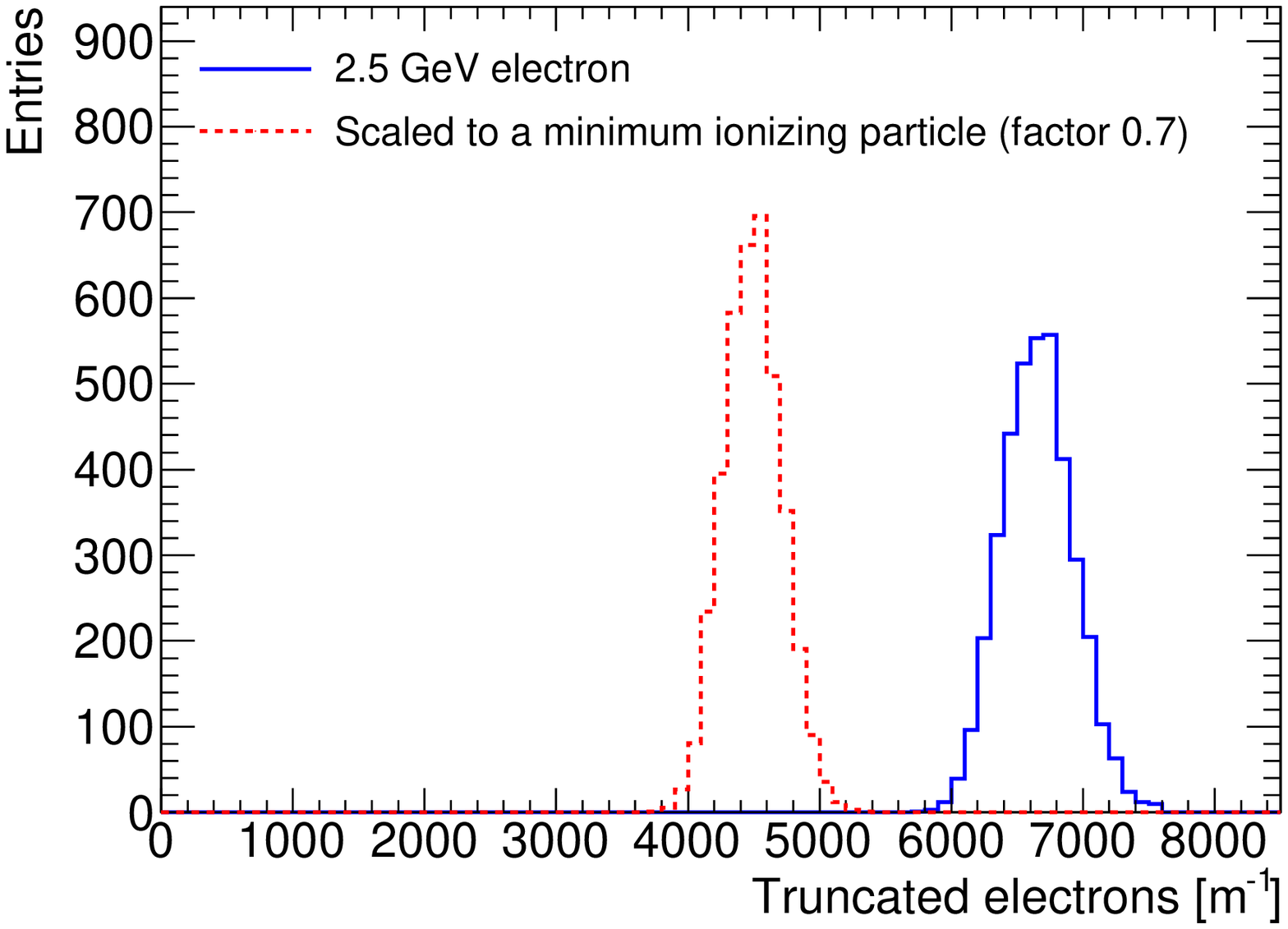}
    \fi
    \caption{Distribution of truncated electrons per meter for the 2.5 GeV electron and the expected distribution for a minimum ionizing particle.}
    \label{fig:dEdx}
\end{figure}

Along the track, the number of electrons is counted for 20 pixel intervals. A fixed fraction of intervals with the highest number of electrons is rejected. Optimally, the top 10\% of intervals with the highest number of electrons are rejected and from the other 90\% of the intervals a truncated sum is obtained. In \autoref{fig:dEdx} the truncated sum is shown for an effective track length of \SI{1}{m} or 83 events. The resolution, expressed as the standard deviation divided by the mean, is 4.1\%.

In order to estimate the separation power, the energy loss distribution for a Minimum Ionizing Particle (MIP) is estimated, see also \autoref{fig:dEdx}. The hit positions of the electron data are scaled track by track by a factor 0.7 to acquire the estimated ionization for a MIP, i.e. \SI{0.7}{m} of electron data is scaled to \SI{1}{m} of expected MIP data. The expected resolution for a \SI{1}{m} MIP track is 4.9\% and the separation between a MIP and a \SI{2.5}{GeV} electron would be 6.2$\sigma$ for a \SI{1}{m} track.

The truncated sum using slices of 20 pixels, does not make full use of the fine granularity of the GridPix detector. We expect that particle identification can be improved by employing the full resolution to resolve primary ionization clusters \cite{Hauschild:2002jh}.

\section{Conclusions and Outlook}
A GridPix detector based on the Timepix3 chip was operated reliably in a test beam setup for the first time. The resolution of the detector in the pixel planes and in the drift direction is limited primarily by diffusion. The additional systematic uncertainties in the pixel plane are smaller than \SI{10}{\um}. Furthermore, by counting the ionization electrons, the energy loss dE/dx can be measured with a precision of 4.1\% for an effective track length of \SI{1}{m}.

The next step towards a TPC for future applications, is the construction of larger size prototype detectors. R\&D has started to build a 4-chip module based on the Timepix3 that can be used to cover larger areas. 
%In the design, all services are possible under the active surface achieving a coverage of \SI{68.9}{\%}.
With these developments, a pixelised readout is on its way to become a mature technology option for a large TPC at a future linear collider.

\section*{Acknowledgements}
This research was funded by the Netherlands Organisation for Scientific Research  NWO.  The authors want to thank the support of the mechanical and electronics departments at Nikhef and the accelerator group at the ELSA facility in Bonn. Their gratitude is extended to the Bonn SiLab group for providing the beam telescope.

\section*{References}

\bibliography{mybibfile}

\end{document}